\documentclass[reprint,amsfonts,amssymb,amsmath,aps,prx,superscriptaddress]{revtex4-1}
\usepackage{times,graphicx,mathptmx}
\usepackage[colorlinks=true,linkcolor=blue,citecolor=blue,urlcolor=blue,bookmarks=false]{hyperref}
\usepackage{CJK}
\usepackage{tikz}

\def\unit#1{\mathord{\thinspace\rm #1}}
\def\func#1{\mathop{\rm #1}\nolimits}%

\begin{document}
\begin{CJK}{UTF8}{bsmi}
\title{Electrostatic Superlattices on Scaled Graphene Lattices}
\author{Szu-Chao Chen (陳思超)}
\affiliation{Department of Physics, National Cheng Kung University, Tainan 70101, Taiwan}
\author{Rainer Kraft}
\affiliation{Institute of Nanotechnology, Karlsruhe Institute of Technology (KIT), D-76021 Karlsruhe, Germany}
\author{Romain Danneau}
\affiliation{Institute of Nanotechnology, Karlsruhe Institute of Technology (KIT), D-76021 Karlsruhe, Germany}
\author{Klaus Richter}
\affiliation{Institut f\"{u}r Theoretische Physik, Universit\"{a}t Regensburg, D-93040 Regensburg, Germany}
\author{Ming-Hao Liu (劉明豪)}
\email{minghao.liu@phys.ncku.edu.tw}
\affiliation{Department of Physics, National Cheng Kung University, Tainan 70101, Taiwan}
\date{\today}

\begin{abstract}

A scalable tight-binding model is applied for large-scale quantum transport calculations in clean graphene subject to electrostatic superlattice potentials, including two types of graphene superlattices: moir\'e patterns due to the stacking of graphene and hexagonal boron nitride (hBN) lattices, and gate-controllable superlattices using a spatially modulated gate capacitance. In the case of graphene/hBN moir\'e superlattices, consistency between our transport simulation and experiment is satisfactory at zero and low magnetic field, but breaks down at high magnetic field due to the adopted simple model Hamiltonian that does not comprise higher-order terms of effective vector potential and Dirac mass terms. In the case of gate-controllable superlattices, no higher-order terms are involved, and the simulations are expected to be numerically exact. Revisiting a recent experiment on graphene subject to a gated square superlattice with periodicity of 35~nm, our simulations show excellent agreement, revealing the emergence of multiple extra Dirac cones at stronger superlattice modulation.

\end{abstract}

\maketitle
\end{CJK}

\section{Introduction}

Graphene, a single layer of carbon atoms arranged in a honeycomb lattice, was first successfully isolated from a single crystal in 2004 \cite{Novoselov2004,Geim2007}, which subsequently triggered further investigations on the intriguing properties of relativistic Dirac fermions \cite{CastroNeto2009}. However, to further uncover more novel electronic properties of the first truly two-dimensional material, the limited mobility of graphene on standard SiO$_2$ substrates turned out to be the main factor restricting mean-free path and phase-coherence length. The discovery of hexagonal boron-nitride (hBN) as an ideal atomically flat substrate for graphene \cite{Dean2010} boosted the development of high-quality graphene devices. Fabrication of devices that involves the encapsulation of graphene between two thin hBN multilayers has become a standard protocol since then \cite{Yankowitz2019}. At the same time, the combination of these two different 2D materials in a so-called van der Waals heterostructure \cite{Geim2013} led to the subsequent discovery of the graphene/hBN moir\'e pattern \cite{Yankowitz2012} arising from the large-scale lattice interference due to the slight lattice constant mismatch. 

At small twist angles, the resulting moir\'e pattern provides a natural source of superlattice potential on graphene with periodicity in the order of $\sim 10\unit{nm}$, leading to the formation of new superlattice minibands in the electronic band structure of graphene at energies reachable by standard electrostatic gating. First experiments revealing new transport phenomena (such as the emergence of the Hofstadter butterfly) were reported in 2013 \cite{Ponomarenko2013,Dean2013,Hunt2013}. In the following years, other exciting transport experiments have been reported \cite{Gorbachev2014,Wang2015,Wang2015a,Lee2016,Handschin2017,KrishnaKumar2017,Chen2017}, as well as a dynamic band structure tuning \cite{Yankowitz2018,Ribeiro-Palau2018}. More recently, another approach for inducing a superlattice potential in graphene has been demonstrated by using patterned dielectrics \cite{Forsythe2018}. Such a gate-tunable superlattice structure allows for the design of arbitrary superlattice geometries with defined periodicity but suffers from technical restrictions.

\begin{figure}[b]
\includegraphics[width=\columnwidth]{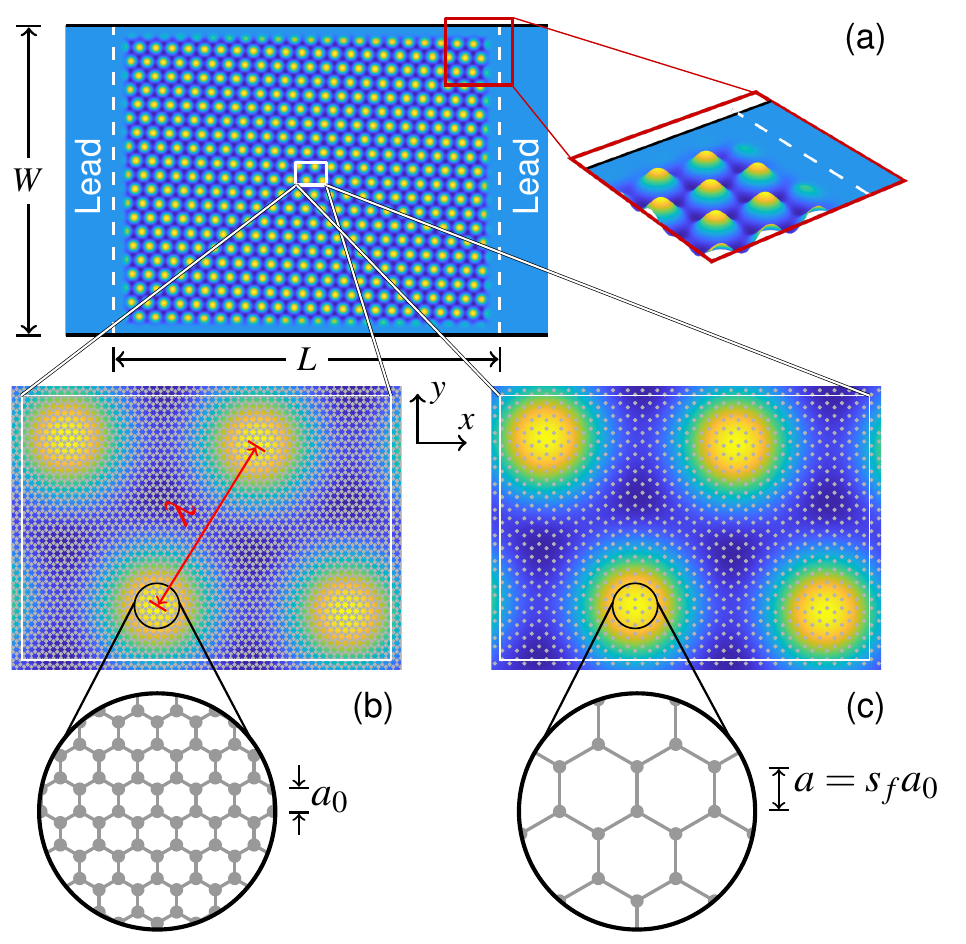}
\caption{(a) A two-terminal graphene device with a superlattice potential within the scattering region only, but not in the attached two leads. The red box is magnified in the surface plot shown at right to indicate the smoothing of the superlattice potential. The white box is magnified in (b) for a genuine graphene lattice and (c) for a scaled graphene lattice.}
\label{fig1}
\end{figure}

On the theory side, most works related to graphene superlattices focus either on calculations for the superlattice-induced mini-band structures \cite{Park2008a,Brey2009,Wu2012,Ortix2012,Kindermann2012,Wallbank2013,Moon2014}, or on predicting transport properties by solving the Dirac equation with oversimplified superlattice model potential \cite{Barbier2010,Burset2011}. On the other hand, quantum transport simulations considering realistic experimental conditions have been relatively rare in the literature \cite{Diez2014,Hu2018}, not to mention a theory work combining quantum transport simulations and mini-band structure calculations, together with transport experiments. This work aims at providing a straightforward method to perform reliable quantum transport simulations, covering both graphene/hBN moir\'e superlattices and gate-induced superlattices. As shown in the following, our transport simulations based on the real-space Green's function method for two-terminal structures as sketched in \autoref{fig1}(a) with the superlattice potential arising either from the graphene/hBN moir\'e pattern or from periodically modulated gating are consistent with transport experiments as well as mini-band structures based on the continuum model. Our method is applicable equally well to multi-terminal structures for simulating, for example, four-probe measurements using the Landauer-B\"uttiker approach \cite{Datta1995}.

This paper is organized as follows. In \autoref{sec methods}, we briefly introduce the theoretical methods used in \autoref{sec results}, including graphene/hBN moir\'e superlattice (\autoref{sec moire superlattice}) and gate-induced superlattice (\autoref{sec gated superlattice}). A summary of this work is provided in \autoref{sec summary}.

\section{Theoretical Approaches}\label{sec methods}

\subsection{Real-space tight-binding model for quantum transport}\label{sec quantum transport}

To perform quantum transport simulations for graphene working in real-space, the scalable tight-binding model \cite{Liu2015} has been proved to be a very convenient numerical trick (see, for example, \cite{Rickhaus2015,Mrenca-Kolasinska2016a,Kolasinski2017,Petrovic2017,Ma2018}): the physics of a graphene system can be captured by a graphene lattice scaled by a factor of $s_f$ such that the lattice spacing is given by $a=s_f a_0$ with $a_0\approx 0.142\unit{nm}$ the carbon-carbon distance and the nearest neighbor hopping is $t = t_0/s_f$ with $t_0\approx 3\unit{eV}$ the hopping parameter for a genuine graphene lattice, as long as the scaled lattice spacing $a$ remains much shorter than all important physical length scales in the graphene system of interest. 

In dealing with graphene superlattices, the newly introduced physical length scale not mentioned in Ref.\ \onlinecite{Liu2015} is the periodicity $\lambda$ of the superlattice. The advantage of the scaling can be easily appreciated by comparing \autoref{fig1}(b) and \autoref{fig1}(c): The former considers a genuine graphene lattice involving lots of carbon atoms, while the latter considers a scaled graphene lattice (here $s_f=2$ for illustrative purposes) involving a much reduced number of lattice sites. As long as $a\ll\lambda$ is satisfied, a reasonably large area covering enough superlattice periods can be implemented in real-space quantum transport simulations to reveal transport properties arising from the superlattice effects.

The model Hamiltonian including the superlattice potential $U_s(x,y)$ using the scaled graphene lattice can be written as
\begin{equation}
H = -t\sum_{\langle i,j\rangle} c_i^\dag c_j + \sum_j U(\mathbf{r}_j)c_j^\dag c_j\ ,
\label{eq tbm}
\end{equation}
where the operator $c_j$ ($c_j^\dag$) annihilates (creates) an electron at site $\mathbf{r}_j=(x_j,y_j)$. The first term in Eq.\ \eqref{eq tbm} represents the clean part of the Hamiltonian which contains nearest neighbor hoppings summing over site indices $i$ and $j$ with $\langle i,j\rangle$ standing for $|\mathbf{r}_i-\mathbf{r}_j|=a$, and the second term is the on-site energy
\begin{equation}
U(x,y) = U_s(x,y)F_s(x,y)+U_0(x,y)\ ,
\label{eq Uxy}
\end{equation}
containing the superlattice potential $U_s(x,y)$ smoothed by a model function $F_s(x,y) = f_s(x,-L/2+d) f_s(-x,L/2-d) f_s(y,-W/2+d) f_s(-y,W/2-d)$ with $f_s(z,z_0) = \{1+\tanh[(z-z_0)/\ell_s]\}/2$, where $\ell_s$ is a smoothing parameter typically taken as $\ell_s=\lambda/4$. The purpose of smearing off the superlattice potential function $U_s$ to zero at a distance $d$ (typically taken as $\lambda$) away from the edges and the leads [see \autoref{fig1}(a) and its inset] is to avoid any spurious effects due to the combination of the superlattice potential and the physical edges of the graphene lattice, as well as to avoid oversized unit cells for the lead self-energies. Any contributions to the on-site energy term other than the superlattice potential are collected in the $U_0$ term in Eq.\ \eqref{eq Uxy}.

With the model Hamiltonian Eq.\ \eqref{eq tbm} constructed, together with self-energies $\Sigma_1$ and $\Sigma_2$ describing the attached two leads (following, for example, Ref.\ \onlinecite{Wimmer2008}), the retarded Green's function at energy $E$ is given by
\begin{equation}
G_r(E) = \frac{\openone}{E-[H+\Sigma_1(E)+\Sigma_2(E)]}\ ,
\label{eq Gr}
\end{equation}
leading to the transmission function
\begin{equation}
T(E) = \func{Tr}[\Gamma_1(E)G_r(E)\Gamma_2(E)G_r^\dag(E)]\ ,
\label{eq T}
\end{equation}
where $\Gamma_j=i(\Sigma_j-\Sigma_j^\dag)$ with $j=1,2$ is the broadening function. In the low-temperature low-bias limit, the conductance across the modeled scattering region is given by the Landauer formula $G = (2e^2/h)T$, where the factor of $2$ accounts for the spin degeneracy. For a pedagogical introduction to the above outlined real-space Green's function, see, for example, Ref.\ \onlinecite{Datta1995}. Note that in most simulations, the full matrix of Eq.\ \eqref{eq Gr} is not needed, suggesting that a partial inversion should be implemented in the numerics to avoid wasting computer memories and CPU time. On the other hand, the matrix version of the Fisher-Lee relation \eqref{eq T} can be implemented as the way it reads.

\subsection{Continuum model for mini-bands and density of states}\label{sec continuum model}

To calculate the mini-band structure of graphene in the presence of a superlattice potential $U_s(\mathbf{r})$, we consider an infinitely large two-dimensional pristine graphene described by $\mathcal{H}_0$ in $k$-space. Following Ref.\ \onlinecite{Park2008a}, we start with the continuum model Hamiltonian near the $K$ valley:
\begin{equation}
\mathcal{H}(\mathbf{k}) = \hbar v_F\begin{pmatrix}
0 & -ik_x-k_y \\
ik_x-k_y & 0
\end{pmatrix}
+U_s(\mathbf{r})\begin{pmatrix}
1 & 0\\
0 & 1
\end{pmatrix}\ ,
\label{eq H continuum}
\end{equation}
where the first term is $\mathcal{H}_0$ and superlattice potential  in the second term is treated as a perturbation. In Eq.\ \eqref{eq H continuum}, the product of the reduced Planck constant $\hbar$ and Fermi velocity $v_F$ is related to the tight-binding parameters through $\hbar v_F = (3/2)ta$, and the two-dimensional wave vector $(k_x,k_y) = \mathbf{k}$ is small relative to the $K$ point. Using the eigenstates of $H_{0}(\mathbf{k})$ as a new basis, we solve the eigenvalue problem of Eq.\ \eqref{eq H continuum} to obtain a set of linear equations:
\begin{equation}
[E-\varepsilon_{s}(\mathbf{k})] c_{s}(\mathbf{k})=\sum_{s',\mathbf{G}}\frac{1+ss'e^{-i\theta_{\mathbf{k},\mathbf{k}-\mathbf{G}}}}{2}\mathcal{U}(\mathbf{G})c_{s'}(\mathbf{k}-\mathbf{G})\ ,
\label{eq linear equations for mini BS}
\end{equation}
where $E$ is the energy eigenvalue of the graphene superlattice, $\varepsilon_{s}(\mathbf{k})=s\hbar v_F k$ is the eigenenergy of $\mathcal{H}_0(\mathbf{k})$ associated with the $s$ branch ($s=1$ for electron above the Dirac point and $s=-1$ for hole below the Dirac point), $\mathcal{U}_{\mathbf{G}}$ is the Fourier component of $U_s(\mathbf{r})$ with the reciprocal lattice vector $\mathbf{G}= m_1\mathbf{G}_1+m_2\mathbf{G}_2$ of the superlattice potential, $\theta_{\mathbf{k},\mathbf{k}-\mathbf{G}}$ is the angle from $\mathbf{k}-\mathbf{G}$ to $\mathbf{k}$, and $c_s(\mathbf{k})$ are the expansion coefficients of the pristine graphene eigenstates.

The infinite-dimensional matrix spanned by the states with wave vectors $\sum_{m_{1},m_{2}} \mathbf{k}+m_{1}\mathbf{G}_{1}+m_{2}\mathbf{G}_{2} $ in Eq.\ \eqref{eq linear equations for mini BS} allows for solving for $E$ and hence calculating the band structure. Since we focus on the low-energy region, a matrix involving states with $|m_{1}| \leq 3$ and $|m_{2}| \leq 3$ is found to be sufficient to attain the convergence of the band structure. 

The density of states $D$ as a function of energy can be calculated by
\begin{equation}
D(\varepsilon )=\frac{2}{(2\pi )^{2}}\int_{1_\mathrm{st}\mathrm{BZ}}\delta
(E-\varepsilon )dk_{x}dk_{y}\ ,
\label{eq D}
\end{equation}
where the integration is taken over the first Brillouin zone. Since $D$ is proportional to the number of energy eigenstates, it can be used to compare with the transport calculations.

\section{Electrostatic superlattices in graphene}\label{sec results}

\subsection{Graphene/hBN moir\'e superlattices}\label{sec moire superlattice}

\begin{figure}[t]
\includegraphics[width=\columnwidth]{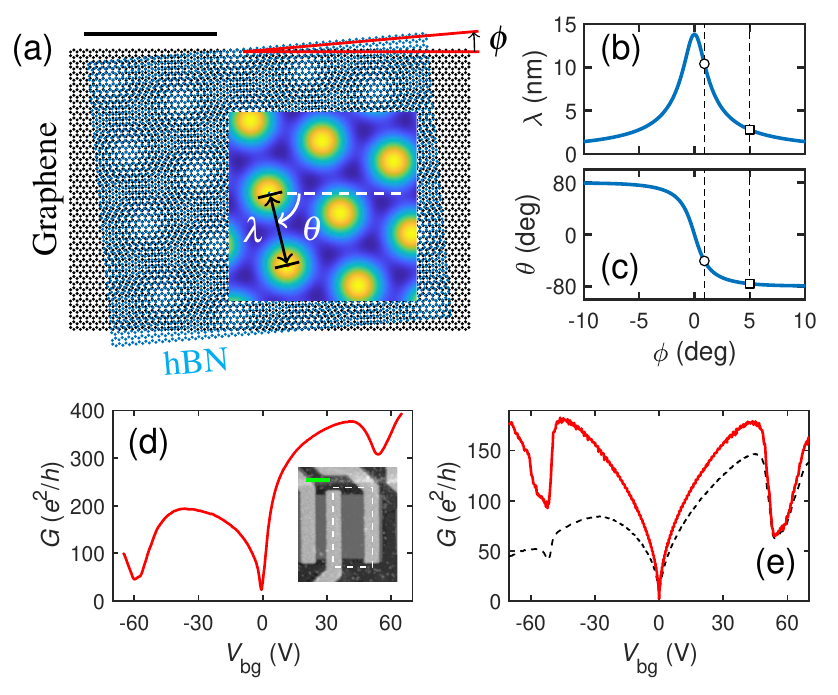}
\caption{(a) An example of stacked graphene and hBN lattices with twist angle $\phi=5^\circ$ showing the resulting moir\'e pattern with wave length $\lambda$ and the orientation angle $\theta$. The overlay of the color map is given by the moir\'e model potential \eqref{eq moire potential} with yellow (bright) and blue (dark) representing the maximum $3V$ and minimum $-3V/2$, respectively. The $\phi$ dependence of $\lambda$ and $\theta$ are shown in (b) and (c), respectively, based on Eq.\ \eqref{eq moire}. (d) Experimentally measured two-terminal conductance as a function of back gate voltage $V_\mathrm{bg}$ at low temperature. Inset: the AFM image of a typical junction similar to the measured device. (e) Simulated conductance as a function of $V_\mathrm{bg}$ at zero temperature with Fermi energy in the leads fixed (black dashed) and floating (red). See the main text for details. Scale bar: $5\unit{nm}$ on (a) and $1\unit{\mu m}$ on the inset of (d).}
\label{fig2}
\end{figure}

Formation of the moir\'e pattern due to the stacking of hBN and graphene lattices has been understood in one of the earliest experiments \cite{Yankowitz2012}. Following their model, the moir\'e pattern results in a triangular periodic scalar potential described by
\begin{equation}
U_s(\mathbf{r}) = V\sum_{j=1,2,3} \cos\left(\mathbf{G}_j\cdot\mathbf{r}\right)\ ,
\label{eq moire potential}
\end{equation}
where $V=0.06\unit{eV}$ is the amplitude of the model potential and $\mathbf{G}_1(\tilde{\lambda},\tilde{\theta})$ is the reciprocal primitive vector of the moir\'e pattern corresponding to the primitive vector $\mathbf{L}_1(\lambda,\theta)=\lambda(\cos\theta,\sin\theta)$ in real space. The orientation angle $\tilde{\theta}$ and wavelength $\tilde{\lambda}$ are related with those in real-space through $\tilde{\theta}=\theta+\pi/2$ and $\tilde{\lambda}=4\pi/\sqrt{3}\lambda$. The other two reciprocal vectors are given by $\mathbf{G}_2(\tilde{\lambda},\tilde{\theta})=\mathbf{G}_1(\tilde{\lambda},\tilde{\theta}+\pi/3)$ and $\mathbf{G}_3(\tilde{\lambda},\tilde{\theta})=\mathbf{G}_1(\tilde{\lambda},\tilde{\theta}+2\pi/3)$. Following \cite{Moon2014} with the zigzag lattice direction arranged along the $x$ axis, the moir\'e wavelength $\lambda$ and the orientation angle $\theta$ of the pattern are given by
\begin{equation}
\begin{aligned}
\lambda &= \frac{1+\epsilon}{\sqrt{\epsilon^2+2(1+\epsilon)(1-\cos\phi)}} a_G \\
\theta &= \arctan\frac{-\sin\phi}{1+\epsilon-\cos\phi}
\end{aligned}\ ,
\label{eq moire}
\end{equation}
where $a_G=\sqrt{3}a_0\approx 0.246\unit{nm}$ is the graphene lattice constant, $\epsilon = (a_\mathrm{hBN}-a_G)/a_G\approx 1.81\%$ is the lattice constant mismatch with $a_\mathrm{hBN}\approx 0.2504\unit{nm}$ the hBN lattice constant, and $\phi$ is the twist angle of the hBN lattice relative to the graphene lattice. An illustrative example with $\phi=5^\circ$ is sketched in Fig.\ \ref{fig2}(a), where an overlay of $U_s(x,y)$ given by Eq.\ \eqref{eq moire potential} is shown to match perfectly the lattice structure of the resulting graphene/hBN moir\'e pattern. For completeness, $\lambda$ and $\theta$ as functions of the twist angle $\phi$ are plotted in \autoref{fig2}(b) and (c), respectively, where the hollow squares mark the $\phi=5^\circ$ example of \autoref{fig2}(a) and the hollow circles mark the $\phi=0.9^\circ$ case corresponding to our transport experiments and simulations to be elaborated below.

According to Ref.\ \onlinecite{Park2008a}, the expected secondary Dirac points in the presence of a triangular superlattice potential occur at the $M$ points of the hexagonal mini-Brillouin zone in $k$ space, with a distance $\tilde{\lambda}/2$ to the $\Gamma$ point (where the normal Dirac point resides). Taking $k = \sqrt{\pi|n|}$ as an estimate for the corresponding carrier density $n$, this suggests that the example of Fig.\ \ref{fig1}(a) with $\phi=5^\circ$ requires a density above $5\times 10^{13}\unit{cm^{-2}}$, which is beyond a reasonable density range from typical electrostatic gating. Indeed, observable graphene/hBN superlattice effects are typically found only in nearly aligned graphene/hBN stacks with a very small twist angle. On the other hand, the $M$ points for the case of $\phi=0.9^\circ$ are expected at a density no more than $4\times 10^{12}\unit{cm^{-2}}$, lying in the typical density range using standard electrostatic gating.

\subsubsection{Transport at zero magnetic field}

To test the validity of the quantum transport simulation illustrated in \autoref{sec quantum transport} using the above moir\'e superlattice model potential Eq.\ \eqref{eq moire potential}, we compare our simulations with the experimental results obtained from a two-terminal device based on a hBN/graphene/hBN stack on a Si/SiO$_2$ substrate, where the crystallographic axis of the graphene flake is aligned with respect to one of the hBN flakes. Electric contact to the graphene is made from the edge of the mesa \cite{Wang2013} with self-aligned Ti/Al electrodes. We use the Si wafer as an overall back gate with a two-layer dielectric consisting of SiO$_2$ with thickness $d_\mathrm{SiO_2} = 300 \unit{nm}$ and the bottom hBN flake with thickness $d_\mathrm{hBN} = 20 \unit{nm}$. A typical exemplary junction similar to the measured device is shown by the atomic-force mircroscope (AFM) image in the inset of Fig.\ \ref{fig2}(d) and marked by the white dashed box. Figure \ref{fig2}(d) shows the two-terminal differential conductance of our sample as a function of the back gate voltage $V_\mathrm{bg}$, measured at low temperature ($\approx 4.1\unit{K}$), using standard low-frequency ($\approx 13\unit{Hz}$) lock-in technique. 

To simulate such a conductance measurement, we have calculated the transmission $T(E)$ as a function of Fermi energy $E$ at zero temperature, and hence the conductance $G(E)=(2e^2/h)T(E)$, based on a $s_f=4$ tight-binding model Hamiltonian, for a two-terminal device similar to \autoref{fig1}(a) with $L=W=500\unit{nm}$, implementing the moir\'e model potential \eqref{eq moire potential} with a twist angle $\phi$. To compare with the experiment on the same voltage axis, we adopt the parallel-plate capacitor formula for the carrier density, $n=(C/e)V_\mathrm{bg}$, and relate $n$ with the Fermi energy through $E=\func{sgn}(n)\hbar v_F\sqrt{\pi|n|}$. These give us
\begin{equation}
V_\mathrm{bg} = \frac{e}{\pi C}\left(\frac{E}{\hbar v_F}\right)^2 \func{sgn}(E)\ ,
\label{eq vbg}
\end{equation}
where $C$ is the back gate capacitance per unit area with $C/e=(\epsilon_0/e)(d_\mathrm{SiO_2}/3.9+d_\mathrm{hBN}/4.2)^{-1}\approx 6.77\times 10^{10}\unit{cm^{-2}V^{-1}}$. Using this capacitance value, the exhibiting conductance dips at $V_\mathrm{bg}=+54.5\unit{V}$ and $-59.9\unit{V}$ observed in \autoref{fig2}(d) correspond to densities about $+3.7\times 10^{12}\unit{cm^{-2}}$ and $-4.0\times 10^{12}\unit{cm^{-2}}$, respectively, so that the twist angle is estimated to lie in the range of $0.85^\circ < \phi < 0.95^\circ$. Indeed, when choosing $\phi=0.9^\circ$ for the moir\'e model potential Eq.\ \eqref{eq moire potential}, the simulated conductance $G(V_\mathrm{bg})$ transformed from $G(E)$ and reported in \autoref{fig2}(e) is found to show excellent agreement with the experiment \autoref{fig2}(d) in the positions of the conductance dips. 

\begin{figure}[t]
\includegraphics[width=\columnwidth]{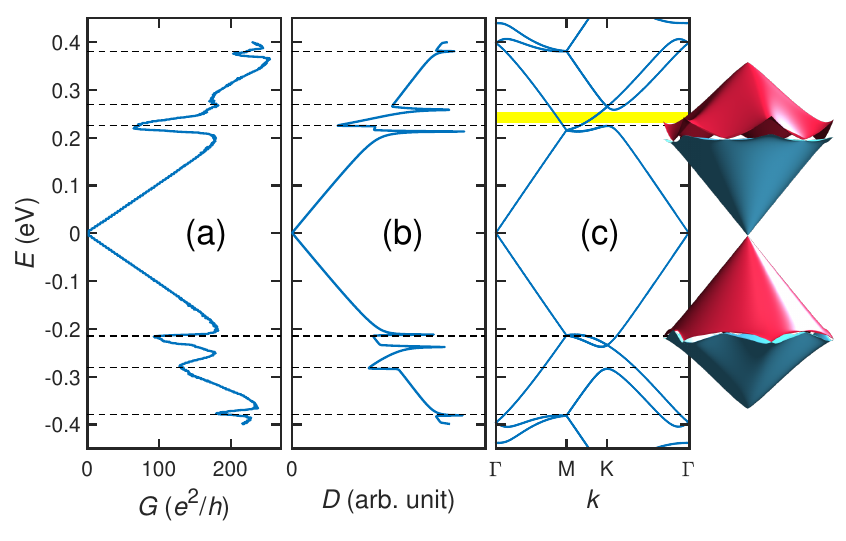}
\caption{Calculated (a) conductance, (b) density of states, and (c) mini-band structures, all based on the same moir\'e superlattice potential \eqref{eq moire potential} with $\phi=0.9^\circ$. In the main panel of (c), an energy window is highlighted in yellow; see the main text. In the side plot of (c), the three-dimensional mini-band structure shows only the lowest two conduction and valence bands.}
\label{fig3}
\end{figure}

Note that the red curve reported in \autoref{fig2}(e) considers leads with on-site and Fermi energies identical to those at the attaching lattice sites; see \autoref{fig1}(a). Compared to \autoref{fig2}(d), the electron-hole asymmetry is less pronounced due to the simple model of Eq.\ \eqref{eq moire potential} from Ref.\ \onlinecite{Yankowitz2012}. Although accounting for an electron doping from the metal leads simply by fixing the Fermi energy in the leads with a positive value can make the conductance curve [black dashed curve in \autoref{fig2}(e) with Fermi energy $0.32\unit{eV}$ in the leads] even more similar to the experiment, the nature of the electron-hole asymmetry observed in \autoref{fig2}(d) comes from higher-order terms such as the effective vector potential and Dirac mass terms \cite{Kindermann2012,Wallbank2013,Moon2014} needed for a model Hamiltonian that can better describe the graphene/hBN moir\'e superlattice. We will continue our discussions with calculations based on leads with ``floating'' Fermi energies.

Without transforming to the gate voltage axis, the original conductance data of \autoref{fig2}(e) as a function of energy is reported in \autoref{fig3}(a) with a wider energy range up to $\pm 0.4\unit{eV}$. Compared to the density of states [\autoref{fig3}(b)] and the band structure [\autoref{fig3}(c)] which are calculated based on the same moir\'e superlattice model potential but within the continuum model (\autoref{sec continuum model}), consistent features in the energy spectrum can be seen. In view of \autoref{fig2}(d)--(e) and \autoref{fig3}, our calculations significantly capture some of the basic properties of the graphene/hBN moir\'e superlattice, at least at zero magnetic field.

\subsubsection{Transport at finite magnetic field}

\begin{figure}[t]
\includegraphics[width=\columnwidth]{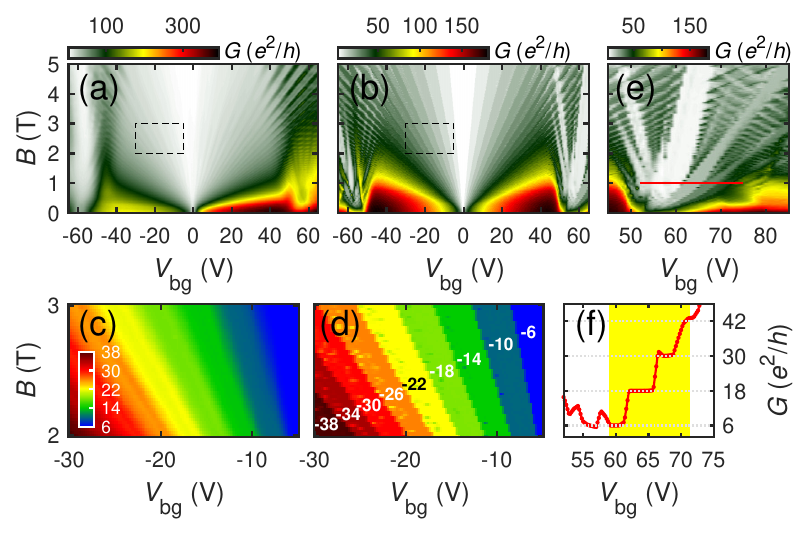}
\caption{Conductance maps $G(V_g,B)$ from (a) experiment and (b) theory, with regions marked by the black dashed boxes magnified and recolored in (c) and (d), respectively. The color bar in (c) is shared with (d), calibrating the conductance in units of $e^2/h$. Numbers on (d) are filling factors on the corresponding plateaus. (e) The simulated conductance map with high gate voltages surrounding the electron-branch secondary Dirac point. The red line corresponds to the horizontal line cut shown in (f).}
\label{fig4}
\end{figure}

We continue our comparison of the experimentally measured and theoretically calculated conductance $G(V_\mathrm{bg})$ with, however, finite magnetic field $B$ perpendicular to the graphene plane, which can be modeled by associating the Peierls phase \cite{Peierls1933,Datta1995} to the hopping $t\rightarrow te^{i\Phi}$ in Eq.\ \eqref{eq tbm}, where $\Phi=(e/\hbar)\int_{\mathbf{r}_j}^{\mathbf{r}_i}\mathbf{A}\cdot d\mathbf{r}$, choosing the Landau gauge $\mathbf{A}=(-yB,0,0)$ for the vector potential $\mathbf{A}$; see the axes shown in \autoref{fig1}. Conductance maps of $G(V_\mathrm{bg},B)$ are reported in \autoref{fig4}(a)/(b) from the experiment/theory showing magnetic field up to $5\unit{T}$ and the gate voltage range same as \autoref{fig2}(d)/(e). Note that the red curves of \autoref{fig2}(d)/(e) correspond exactly to the horizontal line cuts at $B=0$ of \autoref{fig4}(a)/(b). Within the gate voltage range of about $-45\unit{V}\lesssim V_\mathrm{bg} \lesssim +45\unit{V}$, typical relativistic Landau fans can be seen in both experiment and theory. To have a closer look, we magnify the regions marked by the black dashed box on \autoref{fig4}(a)/(b) in \autoref{fig4}(c)/(d) with a different color map to highlight the quantized conductance plateaus. Numbers $-6,-10,\cdots,-38$ on \autoref{fig4}(d) label the filling factor $\nu$ on the corresponding plateau with the expected conductance $|\nu|e^2/h$. Good agreement between experiment [\autoref{fig4}(a) and (c)] and theory [\autoref{fig4}(b) and (d)] within the main Dirac cone can be seen.

At gate voltages $|V_\mathrm{bg}|\gtrsim 45\unit{V}$, transport properties are dominated by the extra Dirac cones arising from the modulating moir\'e superlattice. Discrepancies between the experiment and theory are self-evident. This suggests that the neglected higher-order terms of a more complete model Hamiltonian (see, for example, Refs.\ \onlinecite{Kindermann2012,Wallbank2013,Moon2014}) become important when the magnetic field is strong. Interestingly, we note that in the theory map of \autoref{fig4}(b), some unusual plateaus in the energy range around the electron-branch secondary Dirac point can be observed. We magnify this region in \autoref{fig4}(e) with a horizontal line cut shown in \autoref{fig4}(f), where the gate voltage range showing quantized conductance plateaus is highlighted by a yellow background. 

This $V_\mathrm{bg}$ range, transformed back to the energy through Eq.\ \eqref{eq vbg}, corresponds to an energy window where part of the electron branch of the secondary Dirac cones at $M$ points of the superlattice mini-Brillouin zone are completely isolated. The respective energy window is highlighted also by yellow in \autoref{fig3}(c). Since there are effectively 3 such Dirac cones (six cones on six $M$ points within each mini-Brillouin zone but each cone shared by two neighboring mini-Brillouin zones), the degeneracy factor is expected to be $3\times 2 \times 2 = 12$ with $\times 2$ accounting for spin and another $\times 2$ for valley. Indeed, in the quantum Hall regime, the calculated conductance is quantized to $6,18,30,42$\, $e^2/h$ as shown in \autoref{fig4}(f). Outside this energy (and hence back gate voltage) range, the higher-order Dirac cones are always mixed with background bands, so that no quantized conductance is observed. However, such special energy window leading to the $12$-fold-degeneracy of the Landau levels at the secondary DP is never observed in transport experiments with graphene/hBN moir\'e superlattices \cite{Ponomarenko2013,Dean2013,Lee2016,Handschin2017,KrishnaKumar2017}, including ours shown in \autoref{fig2}(d), indicating once again that the simplified model of Eq.\ \eqref{eq moire potential} containing only the electrostatic scalar potential term is not sufficient to capture transport properties of graphene/hBN moir\'e superlattices at high magnetic fields, i.e. in the quantum Hall regime. As we will see below, when the graphene superlattice potential comes solely from the electrostatic gating, our method becomes exact because in that system no such higher order terms are involved.

\subsection{Gate-controlled superlattices}\label{sec gated superlattice}


To observe any superlattice effects in graphene, the mean free path must be able to cover enough periods of the superlattice potential. This means, either the sample quality must be extraordinary, or the superlattice periodicity must be short enough. When the periodicity is too short, however, the resulting extra Dirac cones appear at too high energy, exceeding the experimentally reachable range. This is why the discovery of the graphene/hBN moir\'e pattern \cite{Yankowitz2012} led to first of its kind studies on graphene superlattices -- the periodicity corresponding to small twist angles turns out to be naturally in a suitable range for experiments; see \autoref{fig2}(b). On the other hand, the superimposed superlattice potential due to the graphene/hBN moir\'e pattern is defined as a hexagonal lattice emerging from the two host lattices.

A more flexible approach to design artificial graphene superlattice structures for band structure engineering was pursued with the realization of electrostatic gating schemes \cite{Forsythe2018}. To create an externally controllable periodic potential, the most intuitive way is to pattern an array of periodic fine metal gates on top of the graphene sample \cite{Dubey2013,Drienovsky2014}. However, due to technical difficulties such as instabilities of nanometer-scale local gates, the low adhesion between metal gates and the inert hBN, etc., such superlattice graphene devices often suffer the problem of very low sample yield \cite{Drienovsky2017}. The basic idea of the new technical breakthrough is to keep the hBN/graphene/hBN sandwich intact, while periodically modulating the gate capacitance. This can be achieved either by using few-layer graphene as a local gate which is subsequently etched with a periodic pattern \cite{Drienovsky2017,Drienovsky2018}, or by etching the dielectric layer with a periodic pattern using a standard uniform back gate underneath the modulated substrate \cite{Forsythe2018}. The latter will be our focus in the rest of this section.

\begin{figure}[t]
\includegraphics[width=\columnwidth]{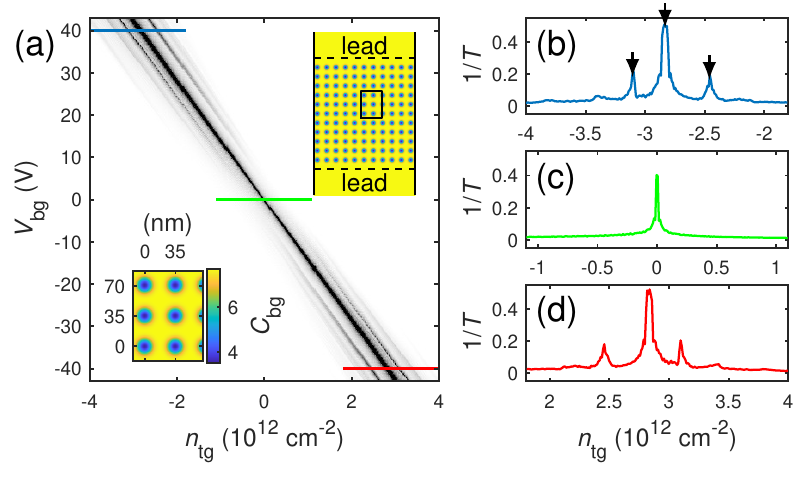}
\caption{(a) Inverse transmission $1/T$ as a function of top-gate-contributed carrier density $n_\mathrm{tg}$ and back gate voltage $V_\mathrm{bg}$ for the simulated virtual device similar to one of those reported in Ref.\ \onlinecite{Forsythe2018}. Upper inset: Device geometry with the modulated back gate capacitance profile showing a spatial modulation of periodicity $\lambda=35\unit{nm}$. The boxed region is magnified in the lower inset with the color bar calibrating the value of back gate capacitance $C_\mathrm{bg}$ in units of $10^{10}\unit{cm^{-2}V^{-1}}$. The three horizontal color lines marked on the main panel of (a) correspond to the line cuts shown in (b)--(d).}
\label{fig5}
\end{figure}

Following the geometry of the device subject to a square superlattice potential with periodicity $\lambda=35\unit{nm}$ presented in Ref.\ \onlinecite{Forsythe2018}, we have performed our own electrostatic simulation to obtain the back gate capacitance showing periodic spatial modulation. We consider an hBN/graphene/hBN sandwich (showing no measurable moir\'e superlattice effects) gated by a global top gate contributing a uniform carrier density $n_\mathrm{tg}$, and a bottom gate at voltage $V_\mathrm{bg}$ with a pre-patterned SiO$_2$ substrate in between. See Fig.\ 1 of Ref.\ \onlinecite{Forsythe2018}. The bottom gate capacitance therefore shows a spatial modulation with a square lattice symmetry, as shown in the lower left inset of \autoref{fig5}(a). 

With the electrostatically simulated position-dependent back gate capacitance per unit area $C_\mathrm{bg}(x,y)$, contributing carrier density $n_\mathrm{bg}(x,y) = [C_\mathrm{bg}(x,y)/e]V_\mathrm{bg}$, together with the uniform $n_\mathrm{tg}$, the resulting superlattice potential is given by $U_s(x,y) = -\func{sgn}[n(x,y)]\hbar v_F\sqrt{\pi|n(x,y)|}$ with $n=n_\mathrm{bg}+n_\mathrm{tg}$, in order to set the global transport Fermi level at zero \cite{Liu2012a}. Slightly different from the case of the graphene/hBN moir\'e superlattice (\autoref{sec moire superlattice}) where the model potential $U_s(x,y)$ given by Eq.\ \eqref{eq moire potential} is independent of the gating, we consider $U(x,y)=U_s(x,y)$ for the on-site energy term \eqref{eq Uxy}, and implement it in the tight-binding Hamiltonian \eqref{eq tbm} with $s_f=6$ to perform quantum transport simulations over a two-terminal structure with $L=420\unit{nm}$ and $W=385\unit{nm}$; see the upper right inset of \autoref{fig5}(a).

To compare with the resistance measurements reported in Ref.\ \onlinecite{Forsythe2018}, we plot the inverse transmission $1/T$ as a function of $n_\mathrm{tg}$ and $V_\mathrm{bg}$ in the main panel of \autoref{fig5}(a), where most areas show high transmission (white regions correspond to low $1/T$). Along the diagonal dark thick line showing high $1/T$ values due to the main Dirac point, multiple satellite peaks can be seen when increasing $|V_\mathrm{bg}|$ and hence the magnitude of the square superlattice potential, signifying the emerging multiple extra Dirac points due to the gate-controlled square superlattice potential. Exemplary line cuts are plotted in \autoref{fig5}(b)--(d) to show clearly the single- and multiple-peak structures, in excellent agreement with the experiment \cite{Forsythe2018}. 

We have also checked the consistency between the calculated mini-band structures and the simulated inverse transmission. Overall, we obtain band structures similar to that reported in Ref.\ \onlinecite{Forsythe2018}, but since each $(n_\mathrm{tg},V_\mathrm{bg})$ point corresponds to a different $U_s(x,y)$ profile and hence a different mini-band structure, an overview consistency-check like in \autoref{fig3} is technically not possible. Instead, the consistency can be checked by comparing the $1/T$ peaks and their corresponding mini-band structure around $E=0$. We have chosen three particular $(n_\mathrm{tg},V_\mathrm{bg})$ configurations corresponding to the three black arrows in \autoref{fig5}(b) marking three of the $1/T$ peaks, at which the $E=0$ Fermi level is expected to hit either the main or the extra Dirac points.

\begin{figure}[t]
\includegraphics[scale=1]{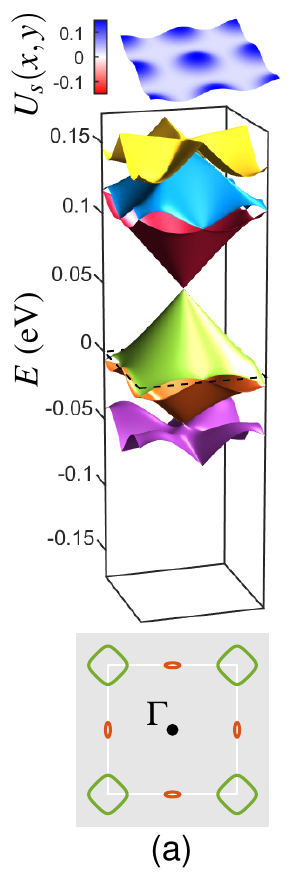}\hfill
\includegraphics[scale=1]{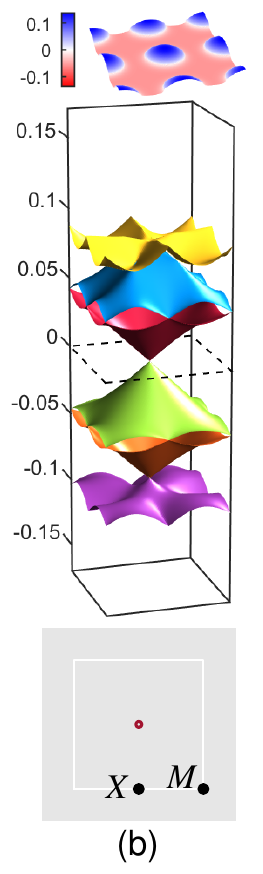}\hfill
\includegraphics[scale=1]{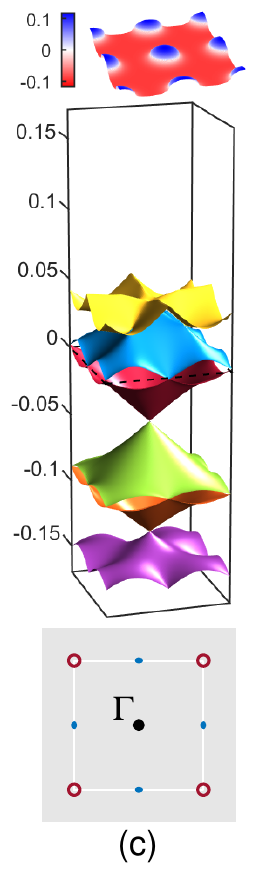}
\caption{Mini-band structures based on the continuum model, showing six bands closest to the main Dirac point, labeled from high to low energy by $C_3$ (yellow), $C_2$ (blue), $C_1$ (red), $V_1$ (green), $V_2$ (orange), and $V_3$ (purple), implementing $U_s(x,y)$ displayed over a range of $2\lambda\times 2\lambda$ above each main panel. Subfigures (a), (b), and (c) correspond to $(n_\mathrm{tg},V_\mathrm{bg})$ configurations labeled by the left, middle, and right black arrows marked in \autoref{fig5}(b), respectively. The black dashed boxes at the middle of the main panels mark the $E=0$ transport Fermi level, where the intersected subband Fermi contours are shown below the corresponding band structure: $V_1$ (green) and $V_2$ (orange) in (a), $C_1$ (red) in (b), and $C_1$ (red) and $C_2$ (blue) in (c). Symmetry points are partly labeled in a way that avoids disturbing the Fermi contours from data.}
\label{fig6}
\end{figure}

These mini-band structures, along with the actual $U_s(x,y)$ profiles implemented individually in the continuum model (\autoref{sec continuum model}) are shown in \autoref{fig6}. Going from low to high $n_\mathrm{tg}$ [left, middle, and right black arrow in \autoref{fig5}(b)], the highest filled energy rises relative to the main Dirac point, corresponding to the sinking of the whole band structure due to our choice of fixing the Fermi level at $E=0$ [\autoref{fig6}(a), (b), and (c)]. As expected, the highest peak in \autoref{fig5}(b) marked by the middle black arrow corresponds to \autoref{fig6}(b), where the main Dirac point is nearly hit; see the lower sub-panel therein and the relevant caption. From the $E=0$ Fermi contours of \autoref{fig6}(a) and (c), the two satellite $1/T$ peaks seen in \autoref{fig5}(b) are mainly contributed by the secondary Dirac points at $X$, labeling the midpoints on the edges of the square mini-Brillouin zone.

Note that the mini-band structures shown in \autoref{fig6}, though corresponding to an increasing uniform $n_\mathrm{tg}$, do not exhibit simply an energy shift without changing the band shape. Compare, for example, the shapes of the lowest subbands. In addition, in this simulated case, no energy windows accommodating completely isolated extra Dirac points can be found. We note, however, that by properly designing the gate capacitance geometry, it is possible to find isolated extra Dirac points, even at $\Gamma$. When isolated extra Dirac points are found, electronic transport is supported solely by the isolated extra Dirac cone, and more novel transport properties of band-engineered graphene superlattices can be explored. This is beyond the scope of the present work and is left as a future direction to further elaborate.

\section{Concluding Remarks}\label{sec summary}

In summary, we have shown that quantum transport simulations based on the scalable tight-binding model \cite{Liu2015} correctly capture transport properties of electrostatic graphene superlattices. In the case of graphene/hBN moir\'e superlattice (\autoref{sec moire superlattice}), the consistency of our simulation and experiment at zero and low magnetic field is rather satisfactory but breaks down at strong magnetic field due to the neglected higher-order terms in a more complete model Hamiltonian \cite{Kindermann2012,Wallbank2013,Moon2014}. In the other case of gated superlattices (\autoref{sec gated superlattice}), without such higher order terms the simulations are expected to be exact. Indeed, compared to the transport experiment with a gate-controlled square superlattice reported in Ref.\ \onlinecite{Forsythe2018}, our simulations show an excellent agreement in revealing the emergence of multiple extra Dirac cones at zero magnetic field. Transport simulations at finite magnetic field for the gated superlattices are expected to reveal also consistent behaviors compared to the experiment, but are left as a future work.

Our work shows that real-space transport simulations based on the scalable tight-binding model \cite{Liu2015} can be extended to treat electrostatic superlattices, whether of the graphene/hBN type or the modulated gate capacitance type, providing consistent results compared to experiments. The method can be immediately applied to take into account, for example, complex local gating or multi-probe transport, in order to make further analysis for transport experiments or even reliable predictions. We note some recent studies working on developing numerical techniques that allow large-scale efficient transport simulations \cite{Beconcini2016,Calogero2018,Papior2019}, but scaling the graphene lattices with an appropriately chosen scaling factor depending on the superlattice periodicity seems to be of least technical complexity and is readily applicable to anyone who is familiar with quantum transport using, for example, real-space Green's function method \cite{Datta1995} or the popular open-source python package KWANT \cite{Groth2014}.

\begin{acknowledgements}

We thank R.\ Huber, J.\ Eroms, and C.\ J.\ Kent for valuable and stimulating discussions. Financial supports from Taiwan Minister of Science and Technology (grant No.\ 107-2112-M-006-004-MY3), Deutsche Forschungsgemeinschaft (projects A07 and Ri681/13-1 under SFB 1277), and Helmholtz Society (program STN) are gratefully acknowledged.

\end{acknowledgements}

\bibliography{../../mhl2}

\end{document}